# Anisotropic exchange interaction of localized conduction-band electrons in semiconductor structures

K.V.Kavokin

*A.F.Ioffe Physico-Technical Institute,*
*194021 Polytechnicheskaya 26, St-Petersburg, Russia*

The spin-orbit interaction in semiconductors is shown to result in an anisotropic contribution into the exchange Hamiltonian of a pair of localized conduction-band electrons. The anisotropic exchange interaction exists in semiconductor structures which are not symmetric with respect to spatial inversion, for instance in bulk zinc-blend semiconductors. The interaction has both symmetric and antisymmetric parts with respect to permutation of spin components. The antisymmetric (Dzyaloshinskii-Moriya) interaction is the strongest one. It contributes significantly into spin relaxation of localized electrons; in particular, it governs low-temperature spin relaxation in n-GaAs with the donor concentration near $10^{16} cm^{-3}$. The interaction must be allowed for in designing spintronic devices, especially spin-based quantum computers, where it may be a major source of decoherence and errors.

**Introduction**

The dynamics of electron spins in semiconductors attracts now a considerable interest due to the idea of using spin for storage, transfer and processing of information (spintronics) [1]. In particular, it has been suggested to use spins of localized electrons in quantum computers [2] either as carriers of quantum information units (qubits) [3], or as agents mediating coupling and coherent transfer of information between qubits realized on nuclear spins [4]. For any spintronic application, strength and symmetry of basic interactions of electron spins are of key importance, because these interactions govern the information transfer as well as spin relaxation and, consequently, decoherence and errors. The spins of two localized electrons are known to be coupled by two kinds of interaction, namely magneto-dipole and exchange interactions. As distinct from the magneto-dipole interaction, the exchange interaction of localized conduction-band electrons is widely believed to be isotropic:

$$\hat{H}_{ex} = 2J\vec{S}_1\vec{S}_2 \qquad (1)$$

where *J* is an exchange constant. Isotropic (scalar) interactions conserve the total spin of the two electrons, and for this reason they do not cause spin relaxation and corresponding information losses in spintronic devices. The isotropic exchange interaction has been supposed to govern the spin structure of the impurity band in n-type semiconductors at low temperature [5].

However, in the crystal environment, the general form of the interaction between two spins-1/2 is more complex:

$$\hat{H}_{ex} = A_{\alpha\beta} S_{1\alpha} S_{2\beta} \qquad (2)$$

where $A_{\alpha\beta}$ is a second-rank tensor defined by the structure symmetry. Anisotropic exchange interactions of this kind, ultimately resulted from the spin-orbit interaction, are well known for paramagnetic ions in crystals [6]. The antisymmetric part of this interaction, known as the Dzyaloshinskii-Moriya interaction [7, 8], is usually written down in a vector form:

$$\hat{H}_{DM} = \vec{d} \cdot \left[\vec{S}_1 \times \vec{S}_2\right] \qquad (3)$$

where the vector $\vec{d}$ is related with the antisymmetric part of the tensor *A*:

$$A_{\alpha\beta} - A_{\beta\alpha} = \varepsilon_{\alpha\beta\gamma} d_\gamma \qquad (4)$$

$\varepsilon_{\alpha\beta\gamma}$ is the third-rank antisymmetric tensor. The Dzyaloshinskii-Moriya interaction can exist when the crystal neighborhood of the two interacting ions lacks inversion symmetry (thus allowing the existence of the vector $\vec{d}$). It arises as the first-order perturbation in the spin-orbit interaction, and is for this reason the strongest anisotropic spin-spin interaction in numerous types of magnetic crystals, including II-VI diluted-magnetic semiconductors [9]. To the best of our knowledge, this interaction has never been considered in the context of localized charge carriers in semiconductors.

In this paper, we argue that the spin-orbit interaction produces an anisotropic part of the exchange interaction between localized conduction-band electrons, having the general form given by Eq.(2), in semiconductor structures that lack inversion symmetry, including practically all low-dimensional structures and also bulk semiconductors with zinc-blend and wurtzite type of the crystal lattice. The main part of the anisotropic exchange interaction has the form of the Dzyaloshinskii-Moriya interaction and may be as strong as several hundredths of the isotropic exchange. This interaction poses considerable problems for the designs of spin-based quantum computers employing electrons bound to natural or artificial localization centers in semiconductor structures, for instance, quantum dots [3] or shallow donors [4]. It can be experimentally detected by its effect upon spin relaxation times in bulk semiconductors and semiconductor nanostructures with appropriate doping.

**Spin-orbit fields in semiconductor structures**

The absence of the spatial inversion in the symmetry group of a semiconductor structure brings about spin-dependent terms in the Hamiltonian of the conduction-band electron, having the following general form:

$$\hat{H}_{SO} = \vec{h}(\vec{k}) \cdot \vec{S} \qquad (5)$$

The vector $\vec{h}$ is an odd function of the electron wave vector $\vec{k}$. In particular, in zinc-blend semiconductors, like GaAs, $\vec{h}$ is cubic in the components of $\vec{k}$ [10, 11]:

$$h_x = \alpha \hbar^3 \left( m_e \sqrt{2 m_e E_g} \right)^{-1} k_x \left( k_y^2 - k_z^2 \right) \qquad (6)$$

where $m_e$ is the effective mass of the electron, $E_g$ is the band gap, $k_x$, $k_y$, $k_z$ are components of the wave vector along cubic axes [100], [010], and [001] respectively. Y and Z-components of $\vec{h}$ are obtained from Eq.(6) by cyclic interchange of indices. The dimensionless coefficient $\alpha$ is equal to 0.07 for GaAs. Dyakonov and Kachorovskii [12] noted that the confinement of the electron envelope wave function in quantum wells creates a considerable mean-squared value of the wave vector component along the structure axis. As a result, $\vec{h}$ becomes linear in the components of the two-dimensional wave vector. For instance, if the structure axis is [001]:

$$h_x = -a k_x; \qquad h_y = a k_y; \qquad h_z = 0 \qquad (7)$$

where $a = \alpha \hbar^3 \left( m_e \sqrt{2 m_e E_g} \right)^{-1} \langle k_z^2 \rangle$. Similar terms exist in bulk semiconductors with wurtzite structure, and in strained zinc-blend crystals [13].

Another contribution to the spin-orbit field is due to gradients of macroscopic potential in the semiconductor crystal. Averaging the spin-orbit interaction over the potential profile in asymmetric quantum wells gives the following dependence of $\vec{h}$ on $\vec{k}$ [14]:

$$h_x = c k_y; \qquad h_y = -c k_x; \qquad h_z = 0 \qquad (8)$$

where $c$ is a constant depending on the shape of the quantum-well potential and properties of the interfaces. This contribution (the so-called Rashba term), as distinct from the spin-orbit terms given by Eqs. (6) and (7), can exist in semiconductors with a centrosymmetric unit cell, like Si and Ge.

The existence of the effective spin-orbit field $\vec{h}$ is well documented experimentally. It causes spin relaxation of electrons via the Dyakonov-Perel' mechanism [11, 13, 12]. It has been also detected directly by passing electric currents through a semiconductor structure: collective movement of the electrons results in a coherent precession of their total spin [15, 16, 17]. Recently, the field $\vec{h}$ has been extensively discussed due to its effect upon weak localization [18].

**Anisotropic exchange: Qualitative consideration**

In localized single-electron states, the odd-in-k terms disappear on averaging over the envelope function of the localized state. However, they do not entirely disappear for the states of two electrons localized at a pair of donors or quantum dots. Instead, they bring forth an anisotropy of the exchange interaction of the electrons. Qualitatively, this can be explained in the following way. When one of the two electrons localized at centers A and B tunnels to the adjacent localization center (say, from A to B), it experiences an influence of the spin-orbit field resulted from the under-barrier motion of the electron. The field causes rotation of the electron's spin through a small angle. Respectively, tunneling of the second electron to the first electron's position (from B to A) is accompanied by the rotation of its spin through the same angle, but in the opposite direction (because $\vec{b}$ changes its polarity for the backward motion). In other words, interchanging the positions of the two electrons goes along with reciprocal rotation of their spins. As a result, there is no way to bring the two electrons into contact, whether at one of the centers or in between, without turning their spins with respect to each other around the direction of the spin-orbit field. One can expect, therefore, that the exchange interaction would couple these tipped spins, $\vec{S}'_A$ and $\vec{S}'_B$, rather then genuine electron-spin operators at centers A and B ($\vec{S}_A$ and $\vec{S}_B$ respectively). This leads to the following heuristic expression for the exchange Hamiltonian

$$\hat{H}_{ex} = 2J\vec{S}'_A \vec{S}'_B = 2J\vec{S}_A\vec{S}_B \cos\gamma + (2J/b^2)(\vec{S}_A\vec{b})(\vec{S}_B\vec{b})(1-\cos\gamma) + (2J/b)(\vec{b}[\vec{S}_A \times \vec{S}_B])\sin\gamma \qquad (9)$$

where γ is the angle of relative rotation of spins, $\vec{b}$ is the effective spin-orbit field that acts upon the electron tunneling from A to B. The first term in Eq.(9) is the usual scalar interaction, the second term corresponds to the symmetric part of the anisotropic exchange, and the third one - to the antisymmetric (Dzyaloshinskii-Moriya) interaction. A rigorous consideration based on the Heitler-London approach, given below, confirms this conclusion and gives the value of the angle γ.

**Formal derivation**

Let us seek the two-electron wave function in the form of a linear combination of functions:

$$\psi_A(1)\psi_B(2)\chi_1^{\sigma_1}\chi_2^{\sigma_2} \quad \text{and} \quad \psi_A(2)\psi_B(1)\chi_1^{\sigma'_1}\chi_2^{\sigma'_2} \qquad (10)$$

where numbers 1,2 numerate electrons, $\psi_A$ and $\psi_B$ are coordinate wave functions localized at centers A and B respectively, $\chi^\sigma$ are spin functions. The functions Eq.(10) approximate eigenfunctions of the full two-electron Hamiltonian $\hat{H}$ at the limit of infinite distance between centers. These functions allow neither for the influence of the potential created by each of the centers upon the one-electron function localized at the other center, nor for spin-orbit interaction. This choice is justified by the smallness of spin-orbit terms in the conduction-band Hamiltonian with respect to the electron binding energy at the center: the influence of the spin-orbit interaction on the wave function of the localized electron is negligible. The functions Eq.10 correspond to degenerate energy levels with $E=E_0$. At finite distances, $\psi_A$ and $\psi_B$ overlap, that results in the appearance of off-diagonal matrix elements of $\hat{H}$ between the two-electron functions of Eq.(10):

$$\begin{aligned}
&\left\langle \psi_A(1)\psi_B(2)\chi_1^{\sigma_1}\chi_2^{\sigma_2} \middle| \hat{H} \middle| \psi_A(2)\psi_B(1)\chi_1^{\sigma'_1}\chi_2^{\sigma'_2} \right\rangle = \\
&= J'\delta_{\sigma_1\sigma'_1}\delta_{\sigma_2\sigma'_2} + \left\langle \psi_A(1)\psi_B(2)\chi_1^{\sigma_1}\chi_2^{\sigma_2} \middle| \hat{H}_{SO} \middle| \psi_A(2)\psi_B(1)\chi_1^{\sigma'_1}\chi_2^{\sigma'_2} \right\rangle = \\
&= J'\delta_{\sigma_1\sigma'_1}\delta_{\sigma_2\sigma'_2} + i\Omega\left\langle \sigma_2 \middle| \vec{b}\vec{S}_2 \middle| \sigma'_2 \right\rangle\delta_{\sigma_1\sigma'_1} - i\Omega\left\langle \sigma_1 \middle| \vec{b}\vec{S}_1 \middle| \sigma'_1 \right\rangle\delta_{\sigma_2\sigma'_2} = \\
&= \left(J'+i\Omega b(S_{2z}-S_{1z})\right)\delta_{\sigma_1\sigma'_1}\delta_{\sigma_2\sigma'_2}
\end{aligned} \qquad (11)$$

where $J'$ is the usual exchange integral (it does not include the spin-orbit interaction), $\Omega$ is the overlap integral of functions $\psi_A$ and $\psi_B$, $\vec{b}=\left\langle \psi_A \middle| \vec{h}(\vec{k}) \middle| \psi_B \right\rangle$, the axis Z is directed along *b*. Note that spin-orbit terms, however small they may be, can not be omitted in calculating matrix elements between degenerate energy levels. The two-electron functions, odd with respect to the permutation of electrons, which are eigenfunctions of the Hamiltonian $\hat{H}$, can then be easily found:

$$\varphi_{IM} = \left( \psi_A(1)\psi_B(2) e^{i\frac{\gamma}{2}(S_{1z}-S_{2z})} + (-1)^I \psi_A(2)\psi_B(1) e^{-i\frac{\gamma}{2}(S_{1z}-S_{2z})} \right) \zeta_{IM} \qquad (12)$$

where $\zeta_{IM}$ is an eigenfunction of the total-spin operator of the two electrons $\vec{I} = \vec{S}_1 + \vec{S}_2$, $M$ is the Z-projection of $I$, $\gamma = \arctan\left(\frac{\Omega b}{J'}\right)$. The corresponding eigenvalues are $E_\pm = E_0 \pm \sqrt{J'^2 + (\Omega b)^2}$. Since we do not include into considerations states with both electrons located near the same center (this is justified due to strong Coulomb repulsion of electrons), the operator of the total spin $\vec{I}$ can be as well represented as $\vec{I} = \vec{S}_A + \vec{S}_B$, where $\vec{S}_A$ and $\vec{S}_B$ are spin operators of electrons localized at centers A and B, respectively. Noting that the first term in the Eq.(12) corresponds to the location of the 1-st and 2-nd electron near centers A and B respectively, and the second term – to the inverted location of electrons, one can substitute the pair of spin operators $\vec{S}_A$ and $\vec{S}_B$ instead of $\vec{S}_1$ and $\vec{S}_2$ in exponents, thus obtaining:

$$\varphi_{IM} = \left( \psi_A(1)\psi_B(2) + (-1)^I \psi_A(2)\psi_B(1) \right) e^{-i\frac{\gamma}{2}(S_{Az}-S_{Bz})} \zeta_{IM} \qquad (13)$$

We come to a two-electron wave function which is a product of an (odd or even) orbital function and a spin function, $\eta_{IM}(\gamma) = e^{-i\frac{\gamma}{2}(S_{Az}-S_{Bz})} \zeta_{IM}$. Functions $\eta_{IM}(\gamma)$ are eigenfunctions of the operator:

$$e^{i\frac{\gamma}{2}(S_{Az}-S_{Bz})} \vec{S}_A \vec{S}_B e^{-i\frac{\gamma}{2}(S_{Az}-S_{Bz})} = \vec{S}'_A \vec{S}'_B \qquad (14)$$

where $\vec{S}'_A$ and $\vec{S}'_B$ are obtained from $\vec{S}_A$ and $\vec{S}_B$ by a rotation around Z through the angles $-\gamma/2$ and $+\gamma/2$ respectively. This immediately yields the expression (Eq.(9)) for the exchange operator $\hat{H}_{ex}$, with $\gamma = \arctan\left(\frac{\Omega b}{J'}\right)$ and $J = \frac{J'}{|J'|}\sqrt{J'^2 + (\Omega b)^2}$.

Let us now find the vector $\vec{b}$ for a few characteristic cases.

i) *A pair of donors in a zinc-blend semiconductor.* In this case $\vec{h}(k)$ is given by Eq.(6). For spherically symmetric functions $\psi(r)$, it is easy to obtain by direct differentiation:

$$\frac{\partial}{\partial z}\left(\frac{\partial^2}{\partial x^2}-\frac{\partial^2}{\partial y^2}\right)\psi(r) = \frac{z(x^2-y^2)}{r^3}\left[\psi'''-3\frac{\psi''}{r}+3\frac{\psi'}{r^2}\right] =$$
$$= \frac{i}{2}\sqrt{\frac{32\pi}{105}}\left(Y_{3,+2}(\theta,\varphi)+Y_{3,-2}(\theta,\varphi)\right)\left[\psi'''-3\frac{\psi''}{r}+3\frac{\psi'}{r^2}\right] \quad (15)$$

where $Y_{3,\pm 2} = -i\sqrt{\frac{105}{32\pi}}\cos^2\theta\sin^2\theta\exp(\pm 2i\varphi)$ is a 3-rd order spherical harmonic, angles $\theta$ and $\varphi$ are defined in the usual way, so that $z=r\cos\theta$, $x=r\sin\theta\cos\varphi$, $y=r\sin\theta\sin\varphi$. Making use of the axial symmetry of the system and of the properties of spherical harmonics [19], one can derive the following expressions for the components of $\vec{b}$:

$$b_z = A_3\int d^3r\,\psi(|\vec{r}-\vec{R}|)\left[\frac{\partial}{\partial z}\left(\frac{\partial^2}{\partial x^2}-\frac{\partial^2}{\partial y^2}\right)\psi(r)\right] =$$
$$= A_3\cos\theta_0\sin^2\theta_0(\cos^2\varphi_0-\sin^2\varphi_0)\int d^3r\left(i\sqrt{\frac{4\pi}{7}}Y_{3,0}\right)\psi(|\vec{r}-\vec{R}|)\left[\psi'''-3\frac{\psi''}{r}+3\frac{\psi'}{r^2}\right] = \quad (16)$$
$$= A_3\frac{R_z(R_x^2-R_y^2)}{R^3}f_3(R)$$

where $f_3(R) = \int d^3r\,\frac{\cos\theta(5\cos^2\theta-3)}{2}\psi(|\vec{r}-\vec{R}|)\left[\psi'''-3\frac{\psi''}{r}+3\frac{\psi'}{r^2}\right]$,

$A_3 = \alpha\hbar^3\left(m_e\sqrt{2m_eE_g}\right)^{-1}$, angles $\theta_0$ and $\varphi_0$ define the direction of $\vec{R}$, so that $R_z=R\cos\theta$, $R_x=R\sin\theta\cos\varphi$, $R_y=R\sin\theta\sin\varphi$. The expressions for other components of $\vec{b}$ are obtained by cyclic interchange of indices.

ii) A similar result for the linear-in-*k* spin-orbit field in a *two-dimensional system* can be obtained with elementary trigonometry:

$$b_x = -A_1\frac{R_x}{R}f_1(R); \quad b_y = -A_1\frac{R_y}{R}f_1(R) \quad (17)$$

for Dyakonov-Kachorovskii terms, and

$$b_x = A_1\frac{R_y}{R}f_1(R); \quad b_y = -A_1\frac{R_x}{R}f_1(R) \quad (18)$$

for Rashba terms. Here $f_1(R) = \int d^2\rho\left[\psi(|\vec{\rho}-\vec{R}|)\frac{\partial\psi(\rho)}{\partial\rho}\cos\theta\right]$, $\vec{\rho}$ is a two-dimentional position vector, $\theta$ is the angle between $\vec{\rho}$ and $\vec{R}$, $A_1$ equals *a* for Dyakonov-Kachorovskii terms (Eq.(7)) and *c* for Rashba terms (Eq.(8)).

**Estimation of the interaction strength**

Asymptotic expressions for the integrals in Eqs.(16), (17), and (18), valid at long distances between centers, are obtained by putting $\cos\theta$ equal to 1 and retaining only terms with the third derivative in Eq.(16). For hydrogen-like centers they read:

$$f_3(R) = \frac{\Omega}{a_B^3}; \qquad f_1 = \frac{\Omega}{2a_B} \qquad (20)$$

where $a_B$ is the effective Bohr radius, $\Omega$ is the overlap integral. The expressions Eq.(20) can be used to estimate $f_1$ and $f_3$ for any other type of potential, substituting in the denominator the corresponding localization radius. Having in mind that $J' \propto \Omega^2 E_B$, where $E_B$ is the electron binding energy at the center, we find that $\gamma$ does not depend on the distance exponentially. The dependence of $\gamma$ on the distance is due to preexponential factors in $\Omega$, $f$, and $J'$, and is rather weak. One can therefore obtain as an order-of-magnitude estimate of angular averages of $\gamma$ at intercenter distances several times greater than the localization radius (when preexponential factors are still close to 1, but asymptotic expressions (Eq.(20)) are already valid):

$$\tan\gamma_3 \approx \frac{A_3}{E_B a_B^3}; \qquad \tan\gamma_1 \approx \frac{A_1}{E_B a_B} \qquad (21)$$

One can see that for the values of parameters typical for most semiconductors these values are much smaller than unity, so that $\tan\gamma \approx \gamma$. Estimates of the typical values of the angle $\gamma$ for a few demonstrative cases are given below:

1) Shallow donors in bulk GaAs: $\gamma \approx 0.01$
2) Donors or quantum dots in a 100A-wide [100] GaAs quantum well: $\gamma \approx 0.1$
3) Donors near interface in Si, with $\vec{R}$ parallel to the surface: $\gamma \approx 0.03$ [20]

To obtain a numerical example of the dependence $\gamma(R)$, a pair of shallow donors in a bulk semiconductor with the zinc-blend lattice is arguably the best model, because localizing potentials, one-particle wave functions, and spin-orbit constants are well known for these systems. Fig.1 displays the results of a numerical calculation of $\gamma$ for donors in GaAs as a function of the distance between donors. The calculation has been performed using exact expressions for the exchange constant $J' = -0.82 E_B \left(\frac{R}{a_B}\right)^{5/2} e^{-2R/a_B}$ [21, 5] and the overlap integral $\Omega = \left(1 + R/a_B + (R/a_B)^3\right) e^{-R/a_B}$ [22] for hydrogen-like centers. Since in this case $\gamma$ depends on the orientation of the donor pair with respect to crystal axes, the angular average of γ(R), $\bar{\gamma}(R) = \left( \frac{1}{4\pi} \int_0^\pi \int_0^{2\pi} \gamma^2(R,\theta,\varphi) \sin\theta \, d\theta \, d\varphi \right)^{1/2}$, and the maximal value of γ(R) corresponding to the orientation of the donor pair along [110], are plotted.

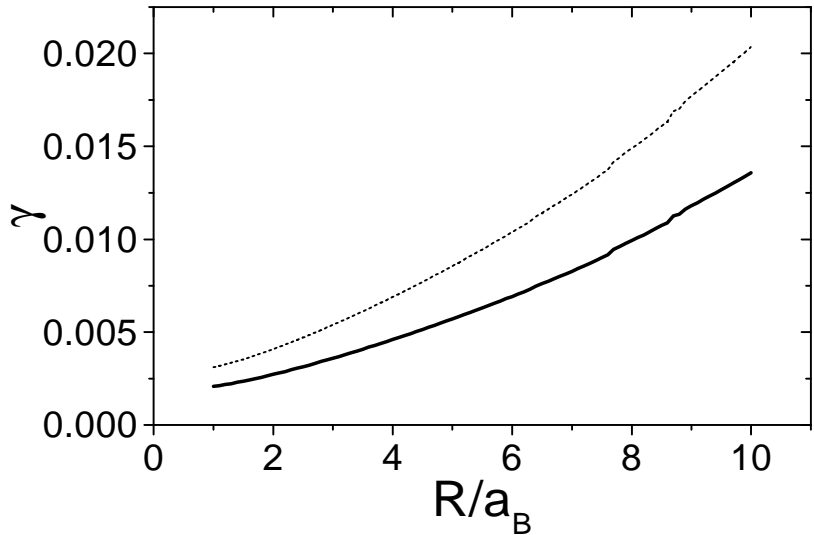

**Fig.1**. Angular average (solid line) and the maximal value of γ (dashed line) for a pair of donors in GaAs vs the distance between donors.

Note that γ rises with $R$. With further increase of $R$ it asymtotically approaches π/2. However, for actual structures at meaningful distances it remains small. This means that, like in the case of $Mn^{2+}$ ions in II-VI semimagnetic semiconductors, the Dzyaloshinskii-Moriya interaction is the strongest and practically only significant part of the anisotropic exchange interaction.

**Comparison with experiment: Spin relaxation in *n*-GaAs.**
A great variety of experimental manifestations of anisotropic spin-spin interactions in systems of magnetic ions or nuclear spins are presently known, including their influence on the magnetic order in magnets [7, 23], spin relaxation [6], selection rules for spin-flip Raman scattering [24, 25] etc. Recently, plenty of optical effects have been observed which are due to anisotropic exchange interaction between electrons and holes in excitons confined in low-dimensional semiconductor structures [26, 27, 28]. Any of those phenomena can be, in principle, considered as a template for designing experiments with localized conduction-band electrons, aimed at revealing the anisotropic exchange interaction, but of course specifics of the energy spectrum, symmetry, and spatial scale of the electron wave function should be taken into account. The use of the conventional [6] or optical [29] EPR, or of the Hanle effect [13], in n-type semiconductors to detect the influence of the anisotropic exchange on the spin relaxation time in the ensemble of donor-bound electrons seems to be the most realistic way. Let us discuss this possibility in more details.

GaAs is the most appropriate model semiconductor for such an inquiry since spin-related phenomena in GaAs have been studied by optical methods for more than 3 decades, and their main features, including spin relaxation mechanisms of free electrons, are well documented [13]. GaAs has the zinc-blend lattice, and therefore the anisotropic exchange interaction should be calculated using Eq.(16). Spin relaxation is caused by random effective fields originated from anisotropic interaction of a given spin with all the other spins in the crystal [6]. Having in mind that the isotropic exchange is by the factor $\gamma^{-1}$ stronger than the anisotropic one, we can apply the dynamic averaging formula for the corresponding spin relaxation time $\tau_{SA}$ (exchange narrowing), yielding:

$$\frac{1}{\tau_{SA}} = \frac{2}{3}\gamma^2 \tau_c^{-1} \qquad (22)$$

where $\tau_c$ is the mean correlation time of the electron spin, governed by flip-flop transitions due to the isotropic part of the exchange interaction. The solid line in Fig.2 shows the calculated $\tau_S$ as a function of donor concentration $n_D$ within the range from $2 \cdot 10^{15}$ cm$^{-3}$ to $2 \cdot 10^{16}$ cm$^{-3}$ (at this latter concentration the Mott transition into the state with metallic conductivity occurs [30]). $\tau_c$ has been calculated by averaging the inverse values of spin splittings of the donor-bound electron, induced by its isotropic exchange interaction with other donors, over the random distribution of donors in the crystal. Exponential dependence of the exchange constant on the average distance between neighboring donors results in very long $\tau_S$ at low donor concentrations, so that other mechanisms of spin decoherence may become competitive. One can suggest: i) thermal activation into the conduction band, where electrons can lose spin orientation by Dyakonov-Perel or Elliot-Yaffet mechanisms; ii) direct interaction with phonons [31]; iii) interaction with lattice nuclei. The latter process should be the most significant at low temperature. The expression for the spin relaxation time of donor-bound electrons due to hyperfine interaction with lattice nuclei was derived by Dyakonov and Perel' [32]. At zero external magnetic field it reads:

$$\frac{1}{\tau_{SN}} = \frac{2}{3}\langle \omega_N^2 \rangle \tau_c \qquad (23)$$

where $\omega_N$ is the frequency of the electron-spin precession in an effective fluctuating magnetic field produced by the nuclear spins within the electron orbit. For shallow donors in GaAs $\omega_N = 5 \cdot 10^8$ c$^{-1}$ [32]. The dashed line in Fig.2 displays the results of calculation of $\tau_{SN}$ performed along Eq.(23) under the assumption that the correlation time $\tau_c$ is governed by the isotropic exchange interaction (this is reasonable at low temperature and low compensation of the semiconductor, when activation into the conduction band and hopping to empty donors are less probable processes than flip-flop transitions). The dotted line gives the spin relaxation time $\tau_S = \left(\tau_{SA}^{-1} + \tau_{SN}^{-1}\right)^{-1}$ which is a result of combined action of the two considered processes.

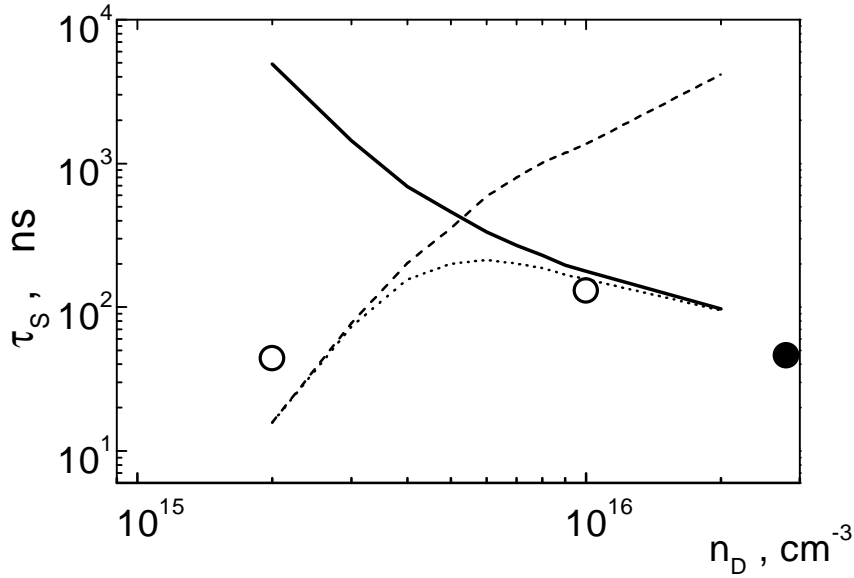

**Fig.2.** Spin relaxation time vs donor concentration in GaAs. Solid line : calculation assuming relaxation exclusively via the anisotropic exchange interaction of donor-bound electrons ($\tau_{SA}$). Dashed line : relaxation via hyperfine interaction with lattice nuclei ($\tau_{SN}$). Dotted line: $\tau_S = \left(\tau_{SA}^{-1} + \tau_{SN}^{-1}\right)^{-1}$. Open circles : experimental data (Refs. [33] and [34]). Solid circle : $\tau_S$ in n-Al$_{0.28}$Ga$_{0.72}$As at 4.2K (determined from experimental data of Ref.[35]).

To the best of our knowledge, two experimental groups have reported measurements of spin relaxation time in GaAs within this concentration range at liquid helium temperatures: Dzhioev et al [33] ($n_D=2\cdot 10^{15}$), and Kikkawa and Awschalom [34] ($n_D =1\cdot 10^{16}$). In both experiments, relaxation times as long as nearly $10^{-7}$ c were measured. The calculated spin relaxation time $\tau_S$ is compared in Fig.2 with the experimental values from Refs. [33] and [34]. For reference, the value of $\tau_S$ for n-Al$_{0.28}$Ga$_{0.72}$As (n=2.8$\cdot 10^{16}$ cm$^{-3}$), determined from experimental data on the Hanle effect [35], is also shown. One can see that the anisotropic exchange interaction is expected to dominate spin relaxation at donor concentrations higher than approximately $7\cdot 10^{15}$cm$^{-3}$, so that the experimental result of Ref [34] can be confidently attributed to the effect considered in this paper. A more detailed theoretical treatment of low-temperature spin relaxation in the impurity band of n-type zinc-blend semiconductors with due allowance for all the mentioned mechanisms will be published elsewhere.

**Implications for quantum computers**

The quantum computer [2] is a hypotetic device that would allow data processing by performing unitary transformations over arrays of two-level quantum systems. The state of each of the two-level systems encodes one quantum information unit, the qubit. It has been proved that general unitary transformations cannot be performed only by manipulating isolated qubits by applying external fields to the two-level systems (one-qubit quantum gates). It is nessesary to perform also two-qubit quantum gates, realized by switching on an interaction between corresponding two-level systems [36]. The operations with qubits should be performed with extreme accuracy. Even with the use of special codes for error correction, large-scale quantum computation would become possible only if the probability of error per quantum gate is less than $10^{-6}$ [37]. There exist several designs of quantum computers exploiting spins of localized electrons in semiconductor structures either as the two-level systems carrying qubits [3], or as the mediator of the interaction between qubits stored on nuclear spins [4]. Spins of localized electrons are attractive for the purposes of quantum computing because they are not subject to the main mechanisms of spin relaxation known for free carriers [13]. For this reason, the hyperfine interaction is considered as the main source of decoherence in quntum-computer cells based on quantum dots [3]. Employing monoisotopic Si with spinless nuclei has been suggested to remove even this channel of spin relaxation [4]. All the designs of spin-based quantum computers rely upon the exchange interaction as the basic means for bringing qubits into contact, besides the exchange interaction is assumed to be isotropic. It follows from the above consideration that this assumption is incorrect for the majority of semiconductor structures (note that the need to manipulate individual qubits makes the designers to place localized electrons near the surface where it would be possible to apply concentrated electric or magnetic fields; as a result, the exchange interaction will be anisotropic even if the host semiconductor is centrosymmetric, like Si). It is evident that, since the anisotropic exchange interaction considered here does not conserve the total spin of the two interacting electrons, it presents an additional source of decoherence. One can easily estimate the probability of the undesirable spin-flip induced by the anisotropic exchange during the swap operation (interchanging directions of two spins by switching on the isotropic exchange for a short period of time), as $p_e \approx \gamma^2$. Since typical values of γ fall into the range from 0.01 to 0.1, the error probability appears to be $10^{-4}$ to $10^{-2}$, which is far beyond the limit of fault tolerant quantum computation, $p_e \leq 10^{-6}$, deduced by Preskill [37]. One could of cource suggest to use the states with a definite spin projection (±1/2) onto the direction of the spin-orbit field $\vec{b}$ as the basic states of the qubit. In this geometry, the anisotropic exchange will not cause spin flips. However, this solution is of limited utility. First, it places constraints on the upscale of quantum-computer cirquits, because in quantum-well or interfacial structures $\vec{b}$ is parallel to the structure plane and depends on the orientation of the pair of localization centers. Therefore, this geometry will not allow two-dimensional arraying of qubits. Then, in quantum dots based on III-V and II-VI semiconductor quantum wells, Dyakonov-Kachorovskii (Eq.7) and Rashba (Eq.8) spin-orbit fields can coexist [38], besides the latter is sensitive to applied electric fields. This may result in changing the direction of $\vec{b}$ with the gate voltage [39], unless the orientation of the pair of quantum dots with respect to crystal axes is carefully chosen. Finally, the Dzyaloshinskii-Moriya interaction does not conserve the squared total spin of the pair of electrons, $I^2$. This means that in addition to errors related to undesirable spin-flips, it will cause phase errors. For example, if the quantization axis is directed along $\vec{b}$, the interactions still mixes states

$\langle 00| = (\langle +1/2|\langle -1/2| - \langle -1/2|\langle +1/2|)/\sqrt{2}$ and

$\langle 10| = (\langle +1/2|\langle -1/2| + \langle -1/2|\langle +1/2|)/\sqrt{2}$, changing the phase between the terms <+1/2|<-1/2| and <-1/2|<+1/2|. Since entangled states like <00| and <10| play a very important role in the theory of quantum computation, the anisotropic exchange interaction may have a serious impact on the operation of the quantum computer. Readout of data from the computer may also be affected. As follows from the above considerations, the spin state corresponding to the symmetric orbital function of the two electrons is not the pure spin singlet <00|. Therefore, the measurement of the spin state of the pair of electrons by checking (with single-electronic techniques) the parity of their orbital wave function, suggested in Ref. [4], will be inevitably accompanied by errors, again with the probability of the order of $\gamma^2 \sim 10^{-4}$-$10^{-2}$.

**Conclusion**

A theoretical study of the exchange interaction between two conduction-band electrons localized at shallow centers (for example, donors or quantum dots) in a semiconductor structure has shown that the interaction may have an anisotropic part governed by the structure symmetry. The anisotropic exchange interaction appears in the effective-mass approximation due to spin-dependent terms in the conduction band Hamiltonian, which are odd in the components of the electron wave vector k. Respectively, the interaction exists in semiconductor structures that lack inversion symmetry, either due to the unit cell geometry (e.g. bulk zinc-blend semiconductors), or as a result of a macroscopic asymmetry of the structure (e.g. asymmetric quantum wells or interfacial layers). The main part of the interaction has the Dzyaloshinskii-Moriya form: $\hat{H}_{DM} = \vec{d} \cdot [\vec{S}_1 \times \vec{S}_2]$, where the direction of vector $\vec{d}$ is governed by the orientation of the pair of localization centers with respect to the crystal axes. The relative strength of the anisotropic interaction with respect to the isotropic exchange interaction weakly depends on the distance between centers and is of the order of a few hundredths. The anisotropic exchange interaction provides an effective channel of spin relaxation in n-GaAs near $n_D = 10^{16}$ cm$^{-3}$. It should be taken into account in anylising spin dynamics of ensembles of localized electrons, which is important for operation of proposed spintronic devices, especially spin-based quantum computers.

**Acknowledgements**

The author is grateful to R.I.Dzhioev, V.L.Korenev, and I.A.Merkulov for very helpful discussions. Partial support of RFBR (Projects 00-15-96756 and 99-02-18082) is acknowledged.

[39] Applying inhomogeneous electric fields with nanometer-sized gates is usually suggested in order to control the overlap of electron wave functions in quantum-computer cells.